\documentclass[aps,amssymb,amsmath, pra]{revtex4}
\usepackage{graphicx}
\usepackage{epsfig}
\usepackage{dcolumn}% Align table columns on decimal point
\usepackage{bm}% bold math
\usepackage{bbm}
\usepackage{hyperref}
\usepackage[up]{subfigure}
\newcommand{\be}{\begin{equation}}
\newcommand{\ee}{\end{equation}}
\newcommand{\bc}{\begin{center}}
\newcommand{\ec}{\end{center}}
\newcommand{\bea}{\begin{eqnarray}}
\newcommand{\eea}{\end{eqnarray}}

\begin{document}
\title{Fractional recurrence in discrete-time quantum walk}
\author{C. M. \surname{Chandrashekar}}
\affiliation{Institute for Quantum Computing, University of Waterloo, 
Ontario N2L 3G1, Canada}
\affiliation{Perimeter Institute for Theoretical Physics, Waterloo, ON, N2L 2Y5, Canada}
%==================================================

\begin{abstract}

%==================================================

Quantum recurrence theorem holds for quantum systems with discrete energy eigenvalues and fails to hold in general for systems with continuous energy. We show that during quantum walk process dominated by interference of amplitude corresponding to different paths fail to satisfy the complete quantum recurrence theorem. Due to the revival of the fractional wave packet, a fractional recurrence characterized using quantum P\'olya number can be seen.
\end{abstract}

%====================================================
\maketitle
\preprint{Version}
%====================================================
\section{Introduction}
\label{intro}
%====================================================
In the dynamics of physical systems, from free particles to stellar dynamics, the recurrence phenomenon have significantly contributed to the better understanding of the systems dynamics \cite{C43}. 
For a classical conservative system, whether discrete or continuous in time, the Poincar\' e recurrence theorem states that any phase-space  configuration $(q,p)$ of a system enclosed in a finite volume will be repeated as accurately as one wishes after a finite interval of time (with no restriction on the interval)\cite{prt}. A similar recurrence theorem is shown to hold in quantum theory as well \cite{BL57}. In a system with a discrete energy eigenvalue spectrum $\{ E_{n} \}$; if $\Psi(t_0)$  is its state vector at the time $t_0$ and $\epsilon$ is any positive number, at least one time $T$ will exist such that the norm of the vector $\Psi(T)-\Psi(t_0)$,
\be
|\Psi(T)-\Psi(t_0)| < \epsilon.
\ee 
Revival of wave packet and its recurrence has attracted considerable attention \cite{R04}, where the dynamics of wave packet in one-dimension has been investigated \cite{Ber96, GRS97}.  Revival of wave packet is in close analogy to the Talbot-effect termed as quantum carpet \cite{BMS01} and Talbot-effect has been experimentally observed in waveguide arrays \cite{IAC05}. 
Revival and recurrence probability has also been studied in both the variants of quantum walk (QW), discrete-time  quantum walk (DTQW ) \cite{ABNV02, SJK08a, SKJ09, Kon09}  and continuous-time quantum walk (CTQW)  \cite{MB05, MB06}. 
 \par
Quantum walks (QWs) are the quantum analog of the dynamics of classical random walks (CRWs) \cite{GVR58, FH65, ADZ93}. The QW  involves superposition of states and moves simultaneously exploring multiple possible paths with the amplitudes corresponding to different  paths interfering with one another. This makes the variance of the QW on a line to grow quadratically with the number of steps compared to the linear growth for the CRW. Due to the speedup, the QW has emerged as a powerful approach to quantum algorithm design \cite{CCD03, SKB03, CG04, AKR05}. QW has also been used to demonstrate the coherent quantum control over atoms and realize quantum phase transition from Mott insulator to superfluid sate and vice versa \cite{CL08}, to explain the phenomenon such as, the breakdown of an electric-field driven system \cite{OKA05} and the direct experimental evidence for wavelike energy transfer within photosynthetic systems\cite{ECR07, MRL08}.   
\par
Many properties of the dynamics of the CRW are very well understood and several analogous properties have been studied in both the variants of the QW.  Recurrence in DTQW differ significantly from the recurrence in CRW. For CRW we can consider the recurrence probability $p_{0}(t)$, that is, the probability of the periodicity of the dynamics that the particle returns to the origin during the time evolution ($t$ steps). It is characterized by the P\'olya number
\be
P_{crw} \equiv 1 - \frac{1}{\sum_{t=0}^\infty p_{0}(t)}.
\ee
If the P\'olya number equals one, the CRW is recurrent, otherwise the walk is transient, i.e., with non-zero probability the particle never returns to the origin. For a CRW to be transient the series $\sum_{t=0}^\infty p_{0}(t)$ must converge \cite{PR90}. P\'olya proved that the one- and two- dimensional CRW are recurrent \cite{GP21}, and for each higher dimension a unique P\'olya number is associated while the  CRWs are transient. 
\par
In standard quantum mechanics, initially localized wave packet in state $|\Psi(0)\rangle$ at $t=0$ which can spread significantly in a closed system can also reform later in the form of a quantum revival in which the spreading reverses itself and the wave packet relocalizes \cite{R04}. The relocalized wave packet can again spread and the periodicity in the dynamics can be seen validating the quantum recurrence theorem. The time evolution of a state $|\Psi(t)\rangle$, or a wave function $\Psi(x,t)$ is given by a deterministic unitary transformation associated with the Hamiltonian. During  quantum evolution we deal with the amplitudes and the probability density $p(x,t) = |\Psi(x,t)|^2$ at position $x$ after time $t$, appears only when we collapse the wave packet to perform measurement. 
\par
Unlike standard wave packet spreading \footnote{During standard wave packet spreading, a wave packet which is initially Gaussian, spreads retaining the Gaussian shape causing increase in the full-width at half maxima (FWHM).}, the QW spreads the wave packet in multiple possible paths with the amplitudes corresponding to different paths interfering. 
\par
In this paper we show that during the evolution of QW, the interference effect and entanglement between the particle and position space delocalizes the wave packet over the position space as small copies of the initial wave packet. These delocalized copies of fractional wave packet fails to satisfy the complete quantum recurrence theorem. However, due to the revival of the fractional wave packets, a fractional recurrence can be seen in QW. The probabilistic characterization of the QW using quantum P\'olya number defined in Ref.\cite{SJK08a} shows  fractional recurrence nature of QW. We also show the exceptional cases of QW that can be constructed by suppressing or minimizing the interference effect and get close to the complete recurrence. 
\par
In Sec. \ref{qw} we briefly describe the standard variants of the discrete-time QW (DTQW) and the continuous-time QW (CTQW). In Sec. \ref{qrecur} we revisit the quantum recurrence theorem.  In Sec. \ref{qwrecur} we discuss the quantum recurrence theorem for QW and show the fractional recurrence nature of the QW on a line and an $n-$cycle before concluding in Sec. \ref{conc}.

%====================================================
\section{Quantum walk}
\label{qw}
%====================================================

The study of QWs has been largely divided into two standard variants: discrete-time QW (DTQW) \cite{ADZ93, DM96, ABN01} and continuous-time QW (CTQW)  \cite{FG98}.  In the CTQW, the walk is defined 
directly on the {\it position} Hilbert space  $\mathcal{H}_p$, whereas, for the 
DTQW it is necessary to introduce an additional {\it coin} Hilbert space $\mathcal{H}_c$, a quantum
coin operation to define the direction in which the particle amplitude
has to evolve. Connection between these two variants and the generic
version of QW has also been studied \cite{FS06, C08}. However, the coin degree of freedom in the DTQW is an advantage over the CTQW as it allows the control of dynamics of the QW \cite{AKR05}. For example, by using a three parameter U(2) or an SU(2) operator as quantum coin operation, the dynamics can be controlled by varying the parameters in the coin operation \cite{CSL08}. Therefore, we take full advantage of the coin degree of freedom and study  the recurrence nature using DTQW.
\par
The DTQW is defined on the Hilbert space $\mathcal  H=  \mathcal H_{c}  \otimes \mathcal H_{p}$ where, $\mathcal H_{c}$ is the {\it coin} Hilbert space and $\mathcal H_{p}$  
is the {\it position} Hilbert space. For a DTQW in one-dimension, the $\mathcal H_{c}$  is spanned  by the basis state of the  particle $|0\rangle$ and  $|1\rangle$ and $\mathcal H_{p}$  is spanned by the basis state of the position $|\psi_{x}\rangle$, $x  \in \mathbb{Z}$.
To implement the DTQW, we will consider a three parameter U(2) operator $C_{\xi, \theta, \zeta}$ of the form
\be 
\label{coin}
C_{\xi,\theta,\zeta}
\equiv    \left(   \begin{array}{clcr}   e^{i\xi}\cos(\theta)    &   &
e^{i\zeta}\sin(\theta)    \\     e^{-i\zeta}    \sin(\theta)    &    &
-e^{-i\xi}\cos(\theta)
\end{array} \right)
\ee
as the quantum coin operation.  The quantum coin operation is applied on the particle state ($C_{\xi,   \theta,  \zeta}  \otimes   {\mathbbm 1}$) when the initial state of the complete system is,
\be
\label{qw:in}
|\Psi_{0}\rangle= (\cos(\delta)|0\rangle + e^{i\eta}\sin(\delta)|1\rangle)\otimes |\psi_{0}\rangle.
\ee
In the above expression $\cos(\delta)|0\rangle + e^{i\eta}\sin(\delta)|1\rangle$ is the state of the particle and $|\psi_{0}\rangle$ is the state of the position. 
The quantum coin operation on the particle is followed by  the conditional unitary shift operation $S$ which delocalizes the wave packet over the position $(x-1)$ and $(x+1)$,
 \be
\label{eq:alter} 
S =\exp(-i\sigma_{z}\otimes \hat{P}l), 
\ee
$\hat{P}$ being  the  momentum operator  and  $\sigma_{z}$, the 
operator corresponding to step of length $l$ depending on the state of the particle 
respectively.  The eigenstates of $\sigma_{z}$ are denoted by $|0\rangle$ and $|1\rangle$. 
Therefore, $S$ can also  be written as,
\begin{eqnarray}
\label{eq:condshift}  S  =  |0\rangle  \langle 0|\otimes  \sum_{x  \in
\mathbb{Z}}|\psi_{x-1}\rangle  \langle \psi_{x} |+|1\rangle  \langle 1
|\otimes \sum_{x \in \mathbb{Z}} |\psi_{x+1}\rangle \langle \psi_{x}|.
\end{eqnarray}
The process of
\be
\label{dtqwev}
 W_{\xi, \theta, \zeta} =
S(C_{\xi,   \theta,  \zeta}  \otimes   {\mathbbm  1})
\ee
is iterated without resorting to the intermediate
measurement to realize large number of steps of the QW. 
$\delta$ and  $\eta$ in Eq. (\ref{qw:in}) can be  varied to  get different
initial state of the particle. The three variable parameters of  the quantum  coin, $\xi$, $\theta$  and $\zeta$  in Eq. (\ref{coin}) can be varied  to  change  the  probability  amplitude  distribution  in  the
position space. By varying the  parameter $\theta$ the variance can be
increased  or  decreased as  a  functional  form, $\sigma^{2}  \approx
(1-\sin(\theta))t^{2}$, where $t$ is the number of steps (iteration).  For a particle with symmetric superposition as initial state the parameter  $\xi$ and $\zeta$ introduces asymmetry in the probability  distribution and their effect on the variance is very small \cite{CSL08}. 
 
%====================================================
\section{Quantum recurrence theorem}
\label{qrecur}
%====================================================
Quantum recurrence theorem in the dynamics of the closed system states that there exist a time $T$  when 
\be
|\Psi(T) - \Psi(t_0) | <  \epsilon ,
\ee
where $\Psi(T) = |\Psi_T\rangle$ is the state of the system after time $T$,  $\Psi(t_0) = |\Psi_0\rangle$ is the initial state of the system and $\epsilon$ is any positive number $\leq$ 2 (both $\Psi(T)$ and $\Psi(t_{0})$ are normalized functions)\cite{BL57}.\\
 The recurrence of complete state of the system or the exact revival happens when all the expectation values of observables $A$ of the two states $|\Psi_T\rangle$ and $|\Psi_0\rangle$ are equal,  that is, 
\be
\langle \Psi_T| A | \Psi_T \rangle = \langle \Psi_0| A | \Psi_0 \rangle.
\ee
\par
In classical dynamics, the characterization of the recurrence nature can be conveniently done using probabilistic measures. Measurements on quantum system leaves the state in one of its basis states with certain probability. Therefore, recurrence in quantum systems can be analyzed using two cases of  comparative evolution of the two identically prepared quantum system with the initial states $|\Psi_{0,0}\rangle$ at position $x = 0$ and time $t=0$. \\
\par
\noindent
{\it Case 1:} Consider two identically prepared particle wave packet which revive completely in position space at time $T$. One of the two particle wave packet at position $x=0$ and time $t=0$ is first evolved to spread in position space and then reverse the spreading till it relocalizes completely at position $x=0$ at time $T$. The measurement performed on this particle will collapse the wave packet at the relocalized position with an expectation value
\be
 \langle \Psi_{0,T}| X | \Psi_{0,T}\rangle =  \langle \Psi_{0,0}| X| \Psi_{0,0} \rangle = 1,
 \ee 
where $X$ is the position operator. After the measurement, the system is further evolved for an additional unit time and the corresponding state can be given by $|\Psi_{x, T_{M}+1}\rangle$, the subscript $M$ stands for the measurement being performed at time $T$. The second  particle wave packet at position $x=0$ and time $t=0$ is evolved up to time $(T+1)$ directly without any measurement being performed at time $T$ and the state can be written as $|\Psi_{x, T+1}\rangle$. Since both the wave packets completely relocalize at position $x=0$ after the evolution for the time period $T$, irrespective of the measurement being performed, the expectation value for both the particle after time $(T+1)$ would be identical, 
\be 
\langle \Psi_{x, T_{M}+1} |X|\Psi_{x, T_{M}+1} \rangle = \langle \Psi_{x, T+1} |X |\Psi_{x, T+1} \rangle
\label{rec}
\ee
with $x$ spanning over all the position space.\\
\par
\noindent
{\it Case 2:} Consider two identically prepared particle wave packet which does not relocalize completely at position $x=0$ at time $T$, that is, revive fractionally or does not revive at all.  
The measurement will collapse the wave packet and return the expectation value 
\be
\label{qwrev}
\langle \Psi_{0,T}| X | \Psi_{0,T} \rangle =\begin{cases}
0   &   ~~ {\rm absence~ of~ revival} \\
1  &  ~~ {\rm fractional~revival}
\end{cases}.
\ee
In this case the two identically prepared particle evolved to time $(T+1)$, one with a measurement being performed at time $T$ and an other particle without any measurement being performed, will not return the same expectation value, that is, 
 \be
 \label{revqw1}
 \langle \Psi_{x, T_{M}+1} |X |\Psi_{x, T_{M}+1} \rangle \neq \langle \Psi_{x, T+1} |X |\Psi_{x,T+1} \rangle
 \ee
at some positions $x$ in the entire position space. If the wave packet does not review at position $x=0$, the left hand side of Eq. (\ref{revqw1}) returns a expectation value 0 for all position $x$ and the right hand side returns 1 for some position $x$ showing the transient nature of the dynamics.  
Therefore in most of the cases, a non-zero values in both the expectation values in the inequality can act as a signature of the fractional recurrence of the quantum state at time $T$ and a zero expectation value on the left hand side shows transient dynamics.  
 \par
From the above analysis we can conclude that, if the system is completely recurrent irrespective of the measurement at time $T$, the two states  $|\Psi_{x, T_M+1}\rangle$ and $|\Psi_{x, T+1}\rangle$ will be equal to one another.  

%====================================================
\section{Fractional recurrence of quantum walk}
\label{qwrecur}
\subsection{On a line}
\label{qwrecurline}
%====================================================
 The state of the particle wave packet after implementing the DTQW of $t$ steps on a line with unit time required to implement each step can be written as 
\be
\sum_{x}|\Psi_{x,t} \rangle = W_{\xi, \theta, \zeta}^t | \Psi_{0, 0}\rangle= \sum_{x}a_x|\Phi_{x,t} \rangle,
\ee
$|\Phi_{x,t} \rangle$ is the state of the delocalized wave packet at each position $x$ in $\mathcal H_{p}$ of any dimension and $|\Psi_{0, 0}\rangle$ is the state of the wave packet before implementing the QW. In the QW process which involves a deterministic unitary evolution, the particle wave packet spreads over the position space forming a small copies of the initial wave packet. During this delocalization process the small copies of the initial wave packet interfere and entangle the position and coin Hilbert space, $\mathcal H_{p}$ and $\mathcal H_{c}$. The interference and the entanglement between the $\mathcal H_{p}$ and $\mathcal H_{c}$ during the standard QW evolution does not permit complete relocalization of the wave packet at initial position after any given number of steps $t$.  Therefore the argument leading to Eq.(\ref{qwrev})  and Eq.(\ref{revqw1}) holds to show that the complete recurrence of the quantum state does not occur during the evolution of the QW process on a line.  However, we note that there is an important conceptual distinction between the full revival of the quantum state and complete relocalization of the DTQW. In the full revival of the quantum state, both the external (position) and internal (particle ) degrees of freedom should be in the same state as it was in the beginning. While for the DTQW,  we are only considering the revival of external  (position) degree of freedom into account.
Therefore we can state :\\
\par
{\it In DTQW evolution on a line and an $n-$cycle dominated by interference of quantum amplitude\footnote{By choosing extreme value of $\theta$ ($0$ or $\pi/2$) in the quantum coin operation, quantum walk on a line can be evolved without constructive or distractive interference taking place.},  there exist no time $T$ where the quantum state of the system revive completely in position degree of freedom and repeat the delocalization and complete revival at regular interval of time (recur)}. \\
\par
 \begin{figure}
\includegraphics[width=9.1cm]{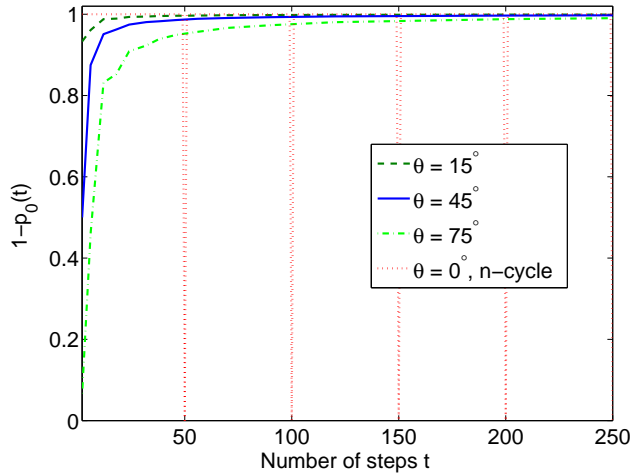}
\caption{The plot of $[1-p_{0}(t)]$,  where $p_{0}(t)$ is the probability of particle at the origin with $t$ being the number of steps in the DTQW evolution.  The plot for QW on a line using different coin operation parameter $\theta = 15^{\circ}, 45^{\circ}$, and $75^{\circ}$ is shown. With increase in $\theta$, the quantum P\'olya number which characterizes the fractional recurrence nature of the QW also increases. The plot with $\theta = 0^{\circ}$ is for a walk on an $n-$cycle with a completely suppressed interference effect, here $n=51$.  For a walk to be completely recurrent, there should exist a $t = T$ where $[1-p_{0}(T)] = 0$, any non-zero value is used to characterize the fractional recurrence nature of the QW and an absolute $1$ for all $t$ would show that the QW is completely transient.}
\label{polyaqw}
\end{figure}
\par
The above statement can also be quantified in the following way:  
After implementing QW, the wave function describing the particle at position $x$ and  time $t$ can be written as a two component vector of amplitudes of particle being at position $x$ at time $t$ with left and right propagating component
\be
\label{qstate}
|\Psi_{x,t}\rangle  =   \left(   \begin{array}{clr}   \Psi^{L} (x, t)   \\     \Psi^{R} (x, t)
\end{array} \right).
\ee
Lets analyze the dynamics of DTQW on wave packet $\Psi(x, t)$ driven by single parameter quantum coin 
\be
\label{qwcoin}
C_{0, \theta, 0} = \left( \begin{array}{ccc}
\cos\theta &~~~& \sin\theta \\
\sin\theta &~~~& - \cos\theta 
\end{array}\right)
\ee
and shift operator $S$ ($W_{0,\theta, 0} = S (C_{0,   \theta,  0}  \otimes   {\mathbbm  1})$).  In terms of left and right propagating component it is given by
\begin{eqnarray}
\label{eq:comp}
\Psi^{L}(x, t+1) &=& \cos\theta \Psi^{L}(x+1,t) +
	 \sin\theta\psi^{R}(x-1,t) \nonumber \\
\Psi^{R}(x, t+1) &=&  \sin\theta \Psi^{L}(x+1, t) -\cos\theta \Psi^{R}(x-1, t). 
\end{eqnarray}
	
Then the probability of wave packet being at position $x$ and $t$ is
\be
\label{prob}
p(x,t) = |  \Psi^{L} (x, t)|^{2} + |\Psi^{R} (x, t)|^{2} .
\ee
and sum of probability over the entire position space is 
\be
\label{probsum}
\sum_{x} p(x, t) = 1.
\ee
After time $t$ with unit time required to implement each step of the quantum walk  on a line will be spread between $x=-t$ to $x=+t$  and the wave function over the position space is  
\be
\sum_{x=-t}^{t} \Psi (x, t)  = \sum_{x=-t}^{t} \left [  \Psi^{L}(x, t) + \Psi^{R}(x, t) \right ].
\ee 
It should be noted that for even number of steps, amplitude at odd labeled positions is $0$ and for odd number of steps amplitude at even labeled position is $0$.  
\par
 For CRW, each step of walk is associated with the {\it randomness} and the {\it probability} of the entire particle therefore, however small the probability is at the origin, it is attributed to the recurrence  of the entire particle. 
Whereas, for the DTQW evolution shown in Eq.  (\ref{eq:comp}) we find that the coin degree of freedom is carried over during the dynamics of the walk making it reversible. The {\it randomness} and the {\it probability} comes into consideration only when the wave packet is collapsed to discard the signature of the coin degree of freedom. Therefore, for a DTQW to be completely recurrent,    the condition
\be
\label{prob1}
p(0,t) = |  \Psi^{L} (0, t)|^{2} + |\Psi^{R} (0, t)|^{2} = 1 .
\ee
has to be satisfied for some time $t$.  Probability at origin after any time $t$, $P(0, t) < 1$ shows the fractional recurrence nature of the QW.  From Eqs. (\ref{eq:comp}) and (\ref{prob}) we can conclude that the Eq. (\ref{prob1}) is satisfied only when $\theta = \pi/2$ and $t$ is even, that is when there is no interference of the quantum amplitudes.  For $\theta = 0$ the two left and right component move in opposite direction without returning and for any $0< \theta < \pi/2$ we get $p(0, t) < 1$ showing the fractional recurrence of the QW.
\par
Alternatively, using quantum Fourier analysis to study the evolution of the DTQW on a line,  it is shown that the amplitude at the origin decreases by  $O(1/\sqrt{t})$. To a very good approximation, independent of position $x$ it has also been shown that the amplitude decrease by $O(1/\sqrt{t})$  ($x$ being the points between the two dominating peaks in the distribution)\cite{ABNV02}. 
Therefore, 
\be
\label{qwexp}
\langle \Psi_{x,t} | \Psi_{x,t} \rangle =  O\left (\frac{1}{t} \right ) < 1.
\ee
and there exists no time $T = t$ where the walk is completely recurrent, whereas QW on a line shows a fractional recurrence nature.   
\par
A probability based characterization of the recurrence nature of the QW, quantum  P\'olya number was defined for an ensemble of identically prepared QW systems by the expression 
\be
\label{qpn}
P_{qw} = 1 - \prod_{t=1}^{\infty} [1- p_{0}(t)]
\ee
where $p_{0}(t)$ is the post measurement recurrence probability of the particle.
Each identically prepared particle is subjected to different number of steps of QW from $1$ to $\infty$ and the probability of the particle at the origin is measured and discarded   \cite{SJK08a}.
The probability that the particle is found at the origin in a single series of such measurement records is the quantum P\'olya number. The quantum P\'olya number was calculated for various coined QWs and it is shown that in the higher dimension it depends both on the initial state and the parameters in the coin operator whereas, for CRW the P\'olya number is uniquely determined by its dimensionality \cite{SJK08b}. 
\par
To show the QW to be completely recurrent adopting a probability based characterization given by Eq. (\ref{qpn}) needs to have at least one of the many particle, each evolved to different steps $t$ from $1$ to $\infty$ to return $p_{0} (t)=1$.
 If $0< p_{0}(t) <1$ for any $t$, then only with certain probability $p_0(t)$ the wave packet  collapses at position $x=0$, that is, prior to measurement the particle existed in superposition of position space. If  $0< p_{0}(t) <1$ for all the particles evolved to different steps from $1$ to $\infty$ then the fractional recurrence of the QW is characterized by the $P_{qw}$.  
\par
In Fig. (\ref{polyaqw}) the plot of $[1-p_{0}(t)]$ is shown for a DTQW on a line where the different coin operation parameter $\theta = 15^{\circ}, 45^{\circ}, 75^{\circ}$.   With increase in $\theta$, the quantum P\'olya number which can also be called as fractional recurrence number, $P_{qw}$ also increases.
\\
%====================================================
\subsection{On an $n-$cycle}
\label{qwrecurcycle}
%====================================================

An $n$-cycle is the simplest finite Cayley graph  with $n$ vertices, number of position in the $\mathcal H_{p}$. This example has most of the features of the walks on the general closed graphs. The unitary shift operation $S$, Eq. (\ref{eq:condshift}) for a QW on an $n-$cycle  is defined by 
\begin{eqnarray}
\label{eq:condshift1}  S^{c}  =  |0\rangle  \langle 0|\otimes  \sum_{x=0}^{n-1}|\psi_{x-1~{\rm mod}~n}\rangle  \langle \psi_{x} |   +|1\rangle  \langle 1
|\otimes \sum_{x=0}^{n-1} |\psi_{x+1~{\rm mod}~n}\rangle \langle \psi_{x}|
\end{eqnarray}
and the quantum state after $t$ steps of QW is written as
\be
|\Psi_{x,t} \rangle = W_{\xi, \theta, \zeta}^t | \Psi_{0, 0}\rangle= \sum_{x = 0}^{n-1} a_x|\Phi_{x,t} \rangle.
\ee
\begin{figure}
\begin{center}
\epsfig{figure=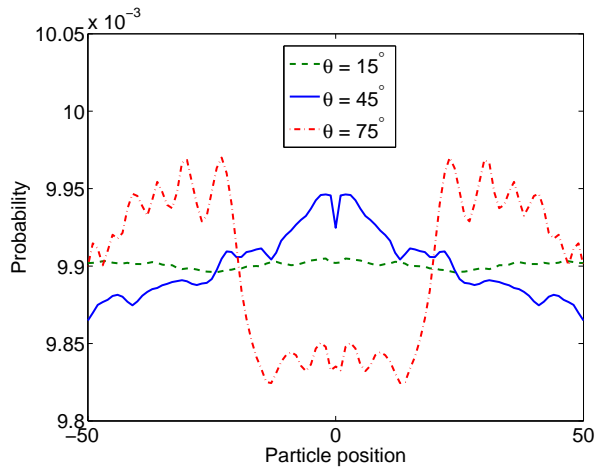, width=8.5cm}
\caption{\label{fig:mixing} Probability distribution of a quantum walker on an $n$-cycle for different value of $\theta$ using coin operation $C_{0,\theta,0}$, where $n$, the number of position, is $101$. The distribution is for 200 cycles. We note that the mixing is faster for lower value of $\theta$. } 
\end{center}
\end{figure}
The  CRW  approaches  a  stationary probability  distribution independent of  its initial  state on a  finite graph.  The time required to approach a stationary distribution is know as mixing time \cite{AAK01}. As we have discussed earlier in case of CRW walk on a line,  a non-zero probability is sufficient to show the recurrence of the entire particle subjected to CRW and stationary distribution returns a non-zero probability.   
\par
Unitary QW, does not converge to any stationary distribution on an $n-$cycle.
But by defining a  time-averaged distribution, 
\be
\overline{p(x,T)} =
\frac{1}{T}  \sum_{t=0}^{T-1}  p(x,  t), 
\ee
\noindent obtained  by  uniformly picking a random time $t$ between $0$ and $(T-1)$, and evolving for $t$ time steps and  measuring to see which vertex it is at, a convergence in the  probability distribution can be  seen even in quantum  case. It has been shown  that the QW on an $n$-cycle mixes in  time $M ={\it O}(n \log n)$, almost quadratically faster than the classical case which is ${\it  O}(n^2)$ \cite{AAK01}.  The mixing time can be optimized on an $n-$ cycle by choosing a lower value of $\theta$ in $W_{0, \theta,  0}$ \cite{CSL08}. Figure  (\ref{fig:mixing})  is the  time  averaged probability distribution of a  quantum walk on an $n$-cycle graph  after $n \log n$ times where $n$ is $101$. It can be seen that the variation of the probability distribution over the position space is least for $\theta = 15^{\circ}$ compared to $\theta = 45^{\circ}$ and $\theta = 75^{\circ}$. 
\par
Uniform distribution in quantum case does not reveal the complete recurrence nature of the particle like it does in classical case. 
\begin{figure}
\begin{center}
\epsfig{figure=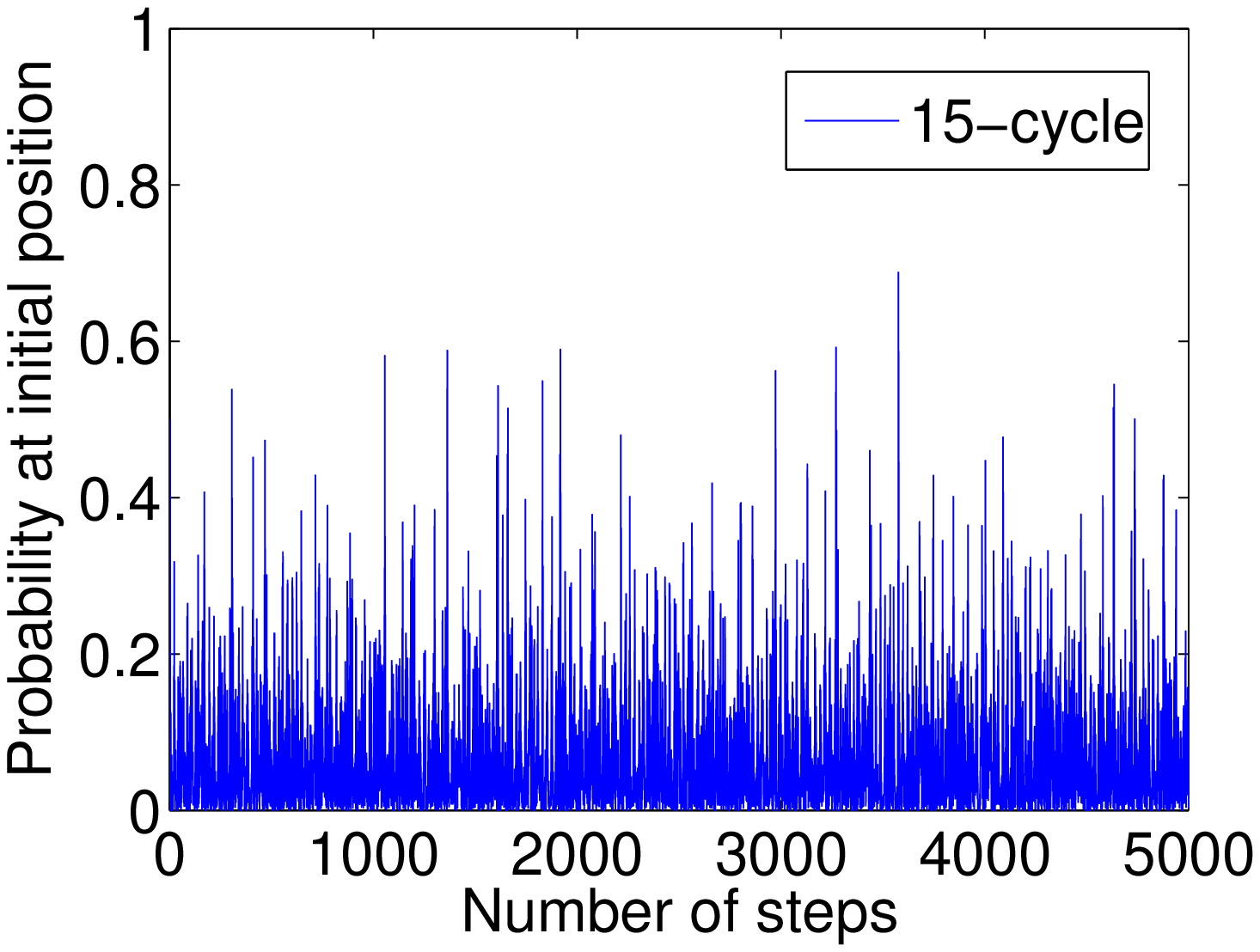, width=7.5cm}
\epsfig{figure=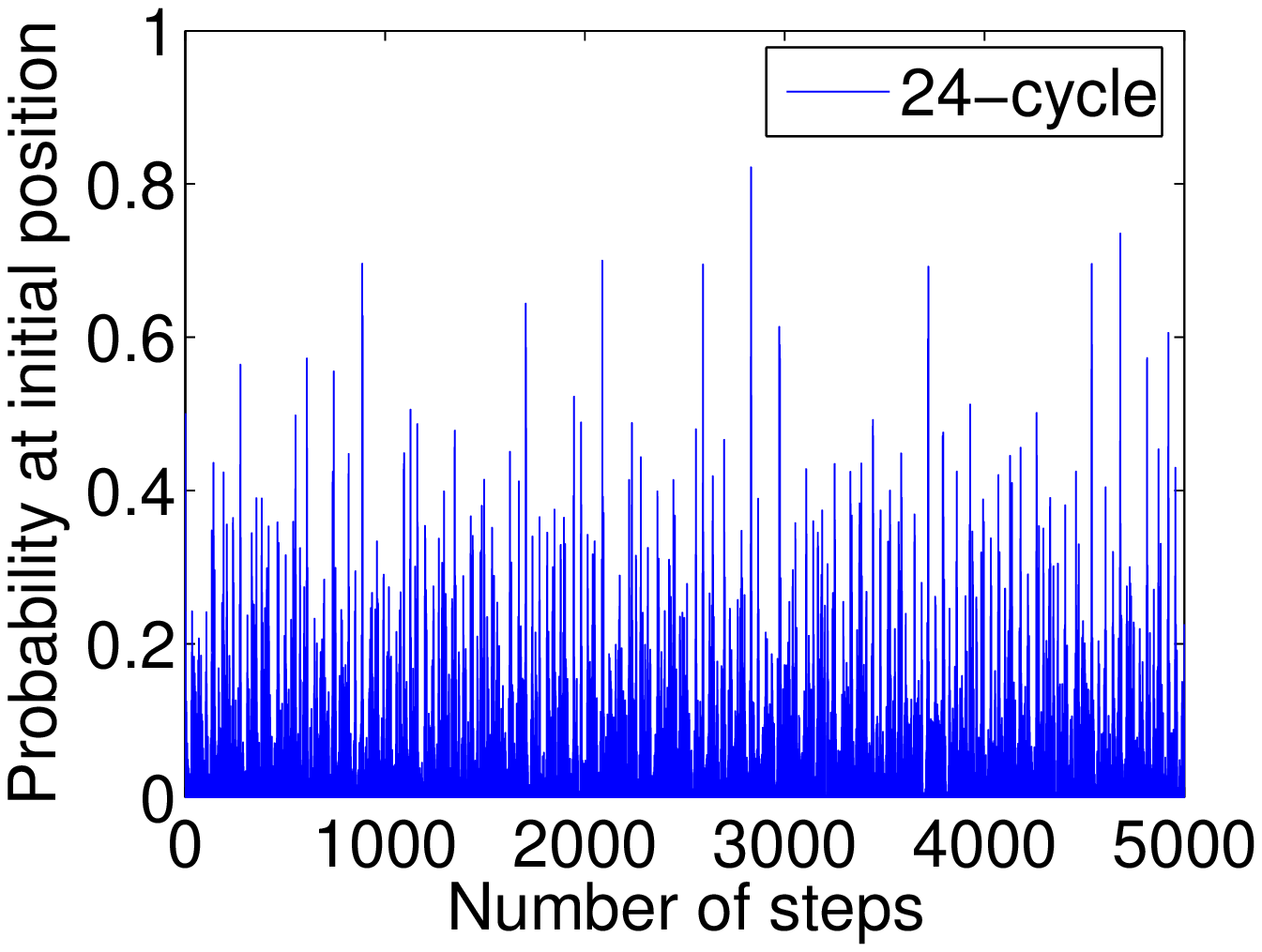, width=7.5cm}\\
~~~~~~~(a)~~~~~~~~~~~~~~~~~~~~~~~~~~~~~~~~~~~~~~~~~~~~~~~~~~~~~~~~~~~~~~~~~(b)
\caption{\label{fig:fracrev1} Probability at the initial position after different number of steps of quantum walk on $15-$ and $24-$ cycle.  This clearly shows that even for steps as large as 5000, there is no signature of complete recurrence. The distribution is obtained using Hadamard operation, $C_{0, \pi/4, 0}$ as quantum coin operation. }
\end{center}
\end{figure}
\begin{figure}
\begin{center}
\epsfig{figure=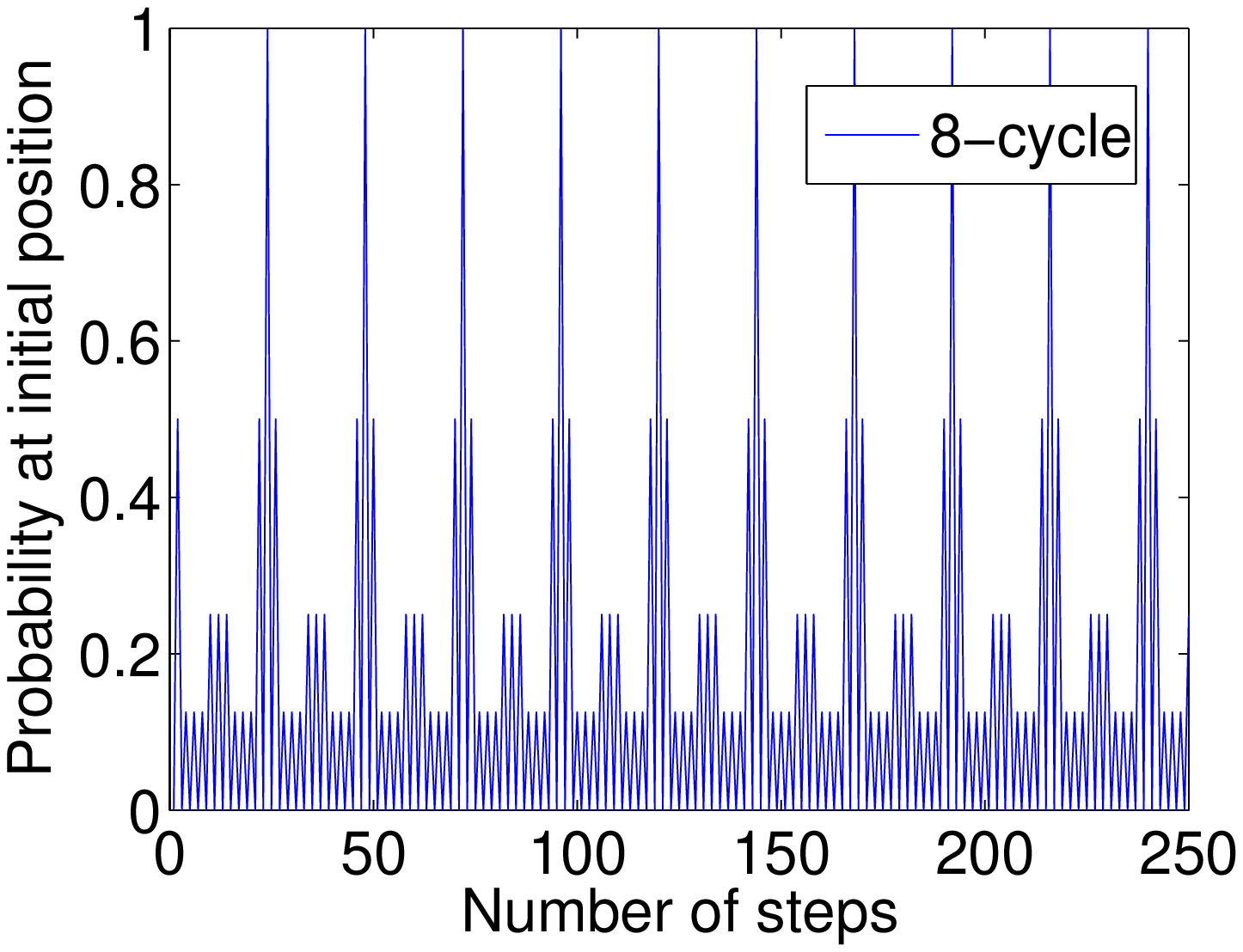, width=7.1cm}
\epsfig{figure=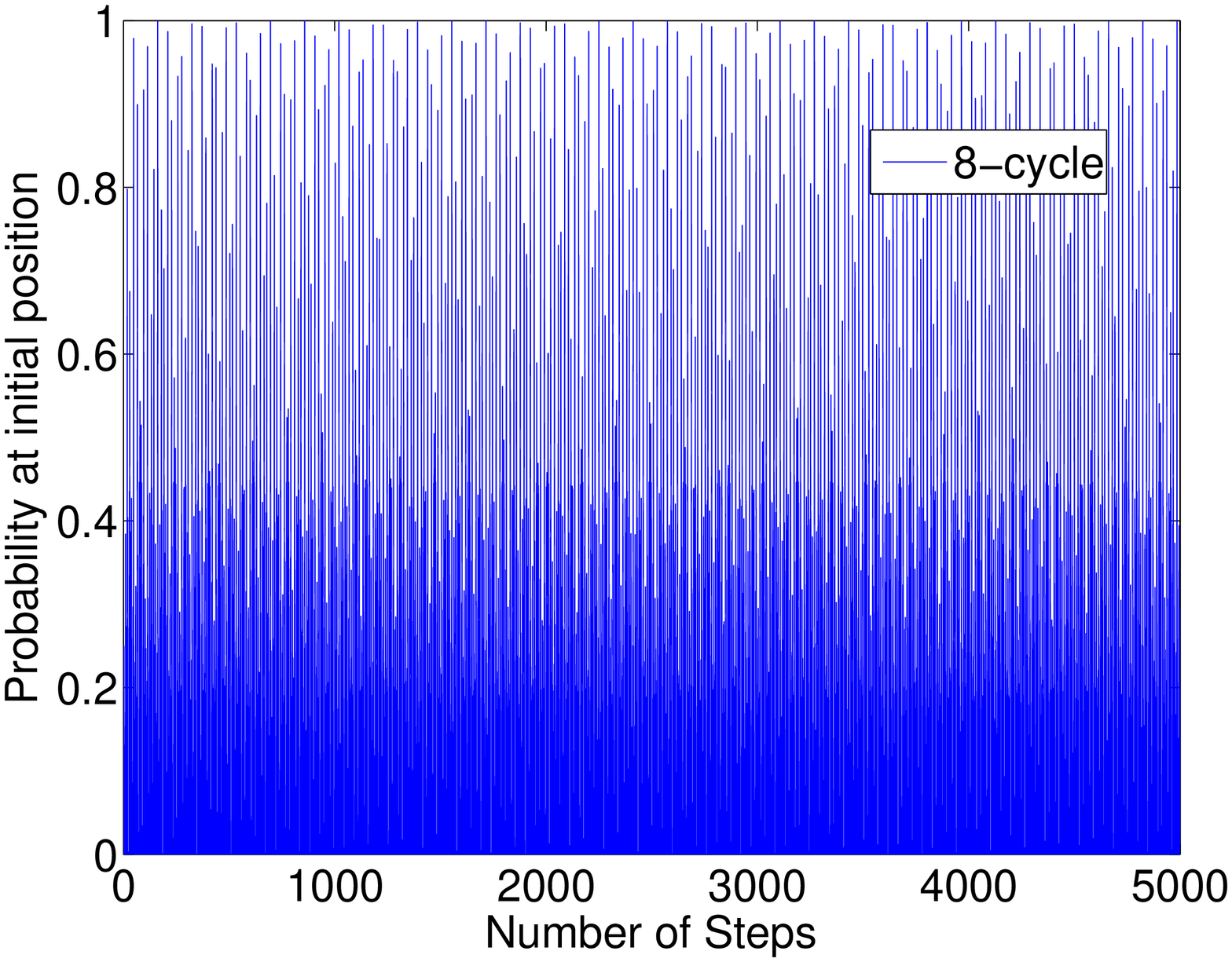, width=7.1cm}\\
~~~~~~~(a)~~~~~~~~~~~~~~~~~~~~~~~~~~~~~~~~~~~~~~~~~~~~~~~~~~~~~~~~~~~~~~~~~(b)
\caption{\label{fig:rev8cycle} Probability at the initial position on $8-$cycle. (a)
The distribution is obtained using Hadamard coin operation $C_{0, 45^{\circ}, 0}$, due to return of amplitudes to initial position (constructive interference at the origin) before interference of amplitude dominates uniformly over the entire vertices, recurrence is seen.  (b) The use of quantum coin, $C_{0, 30^{\circ}, 0}$ during the evolution does not lead to complete constructive interference at the origin and hence complete recurrence is not seen.} 
\end{center}
\end{figure}
\begin{figure}
\begin{center}
\epsfig{figure=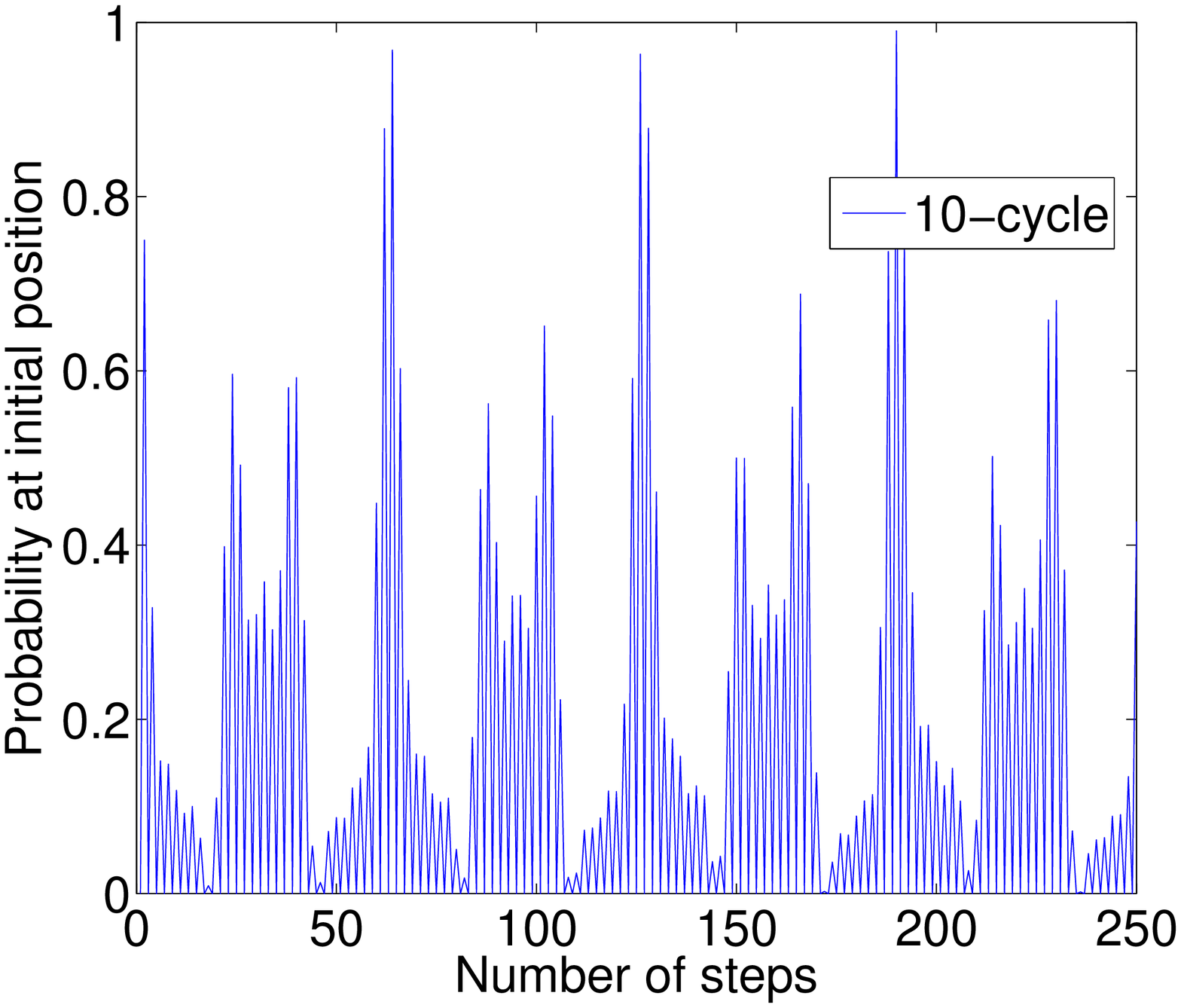, width=7.1cm}
\epsfig{figure=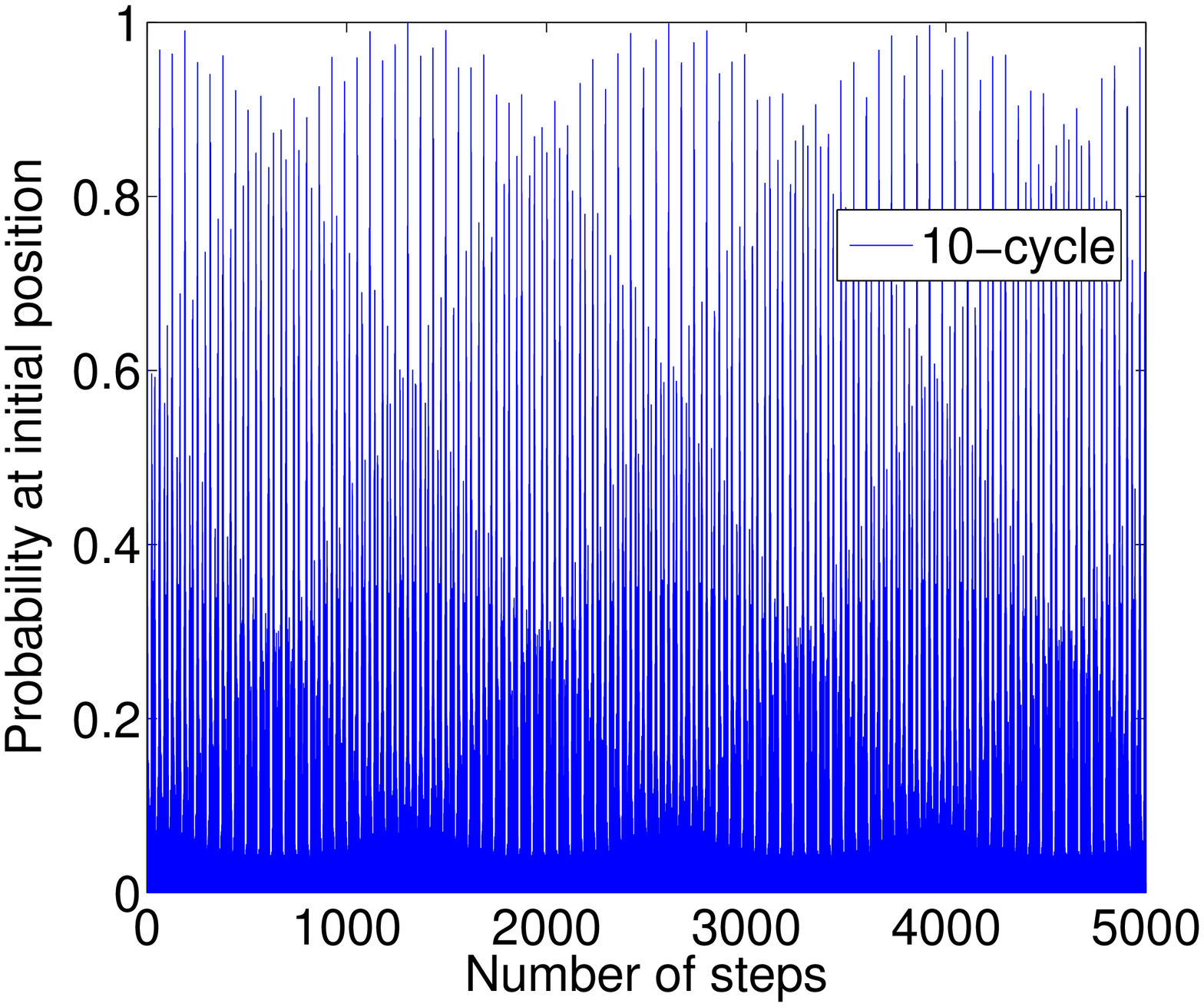, width=7.1cm}\\
~~~~~~~(a)~~~~~~~~~~~~~~~~~~~~~~~~~~~~~~~~~~~~~~~~~~~~~~~~~~~~~~~~~~~~~~~~~(b)\\
\epsfig{figure=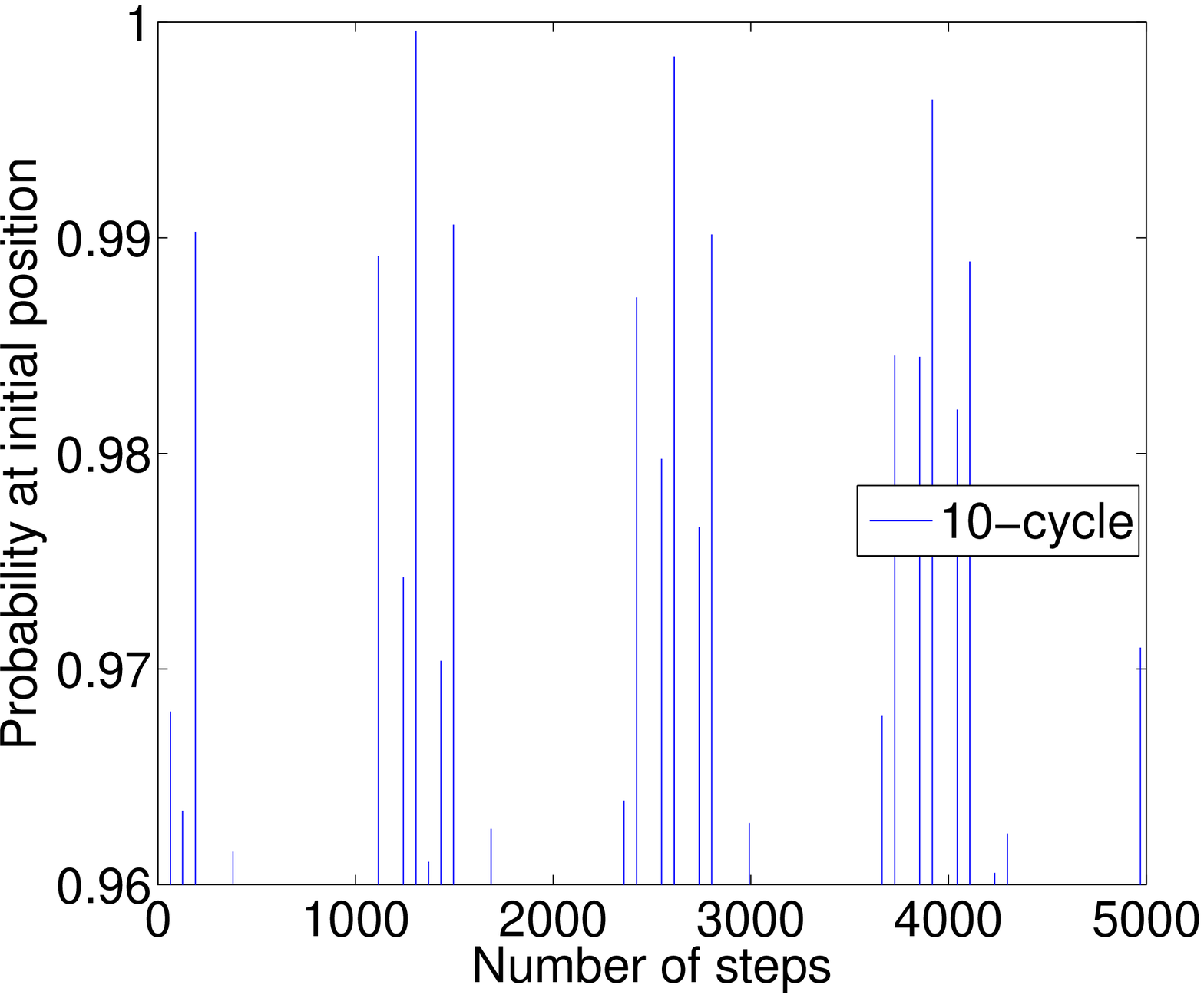, width=7.1cm}\\
(c)
\caption{\label{fig:rev10cycle} Probability at the initial position on $10-$cycle. The distribution is obtained using Hadamard operation $C_{0, 45^{\circ}, 0}$. A small deviation from the complete recurrence can be seen. (a) and (b) are probability at initial position for quantum walk up to 250 steps and 5000 steps respectively and (c) is the close up of the probability and we note that the probability is not exact 1 at any time within 5000 steps.} 
\end{center}
\end{figure}
\begin{widetext}
\begin{figure}
\begin{center}
\epsfig{figure=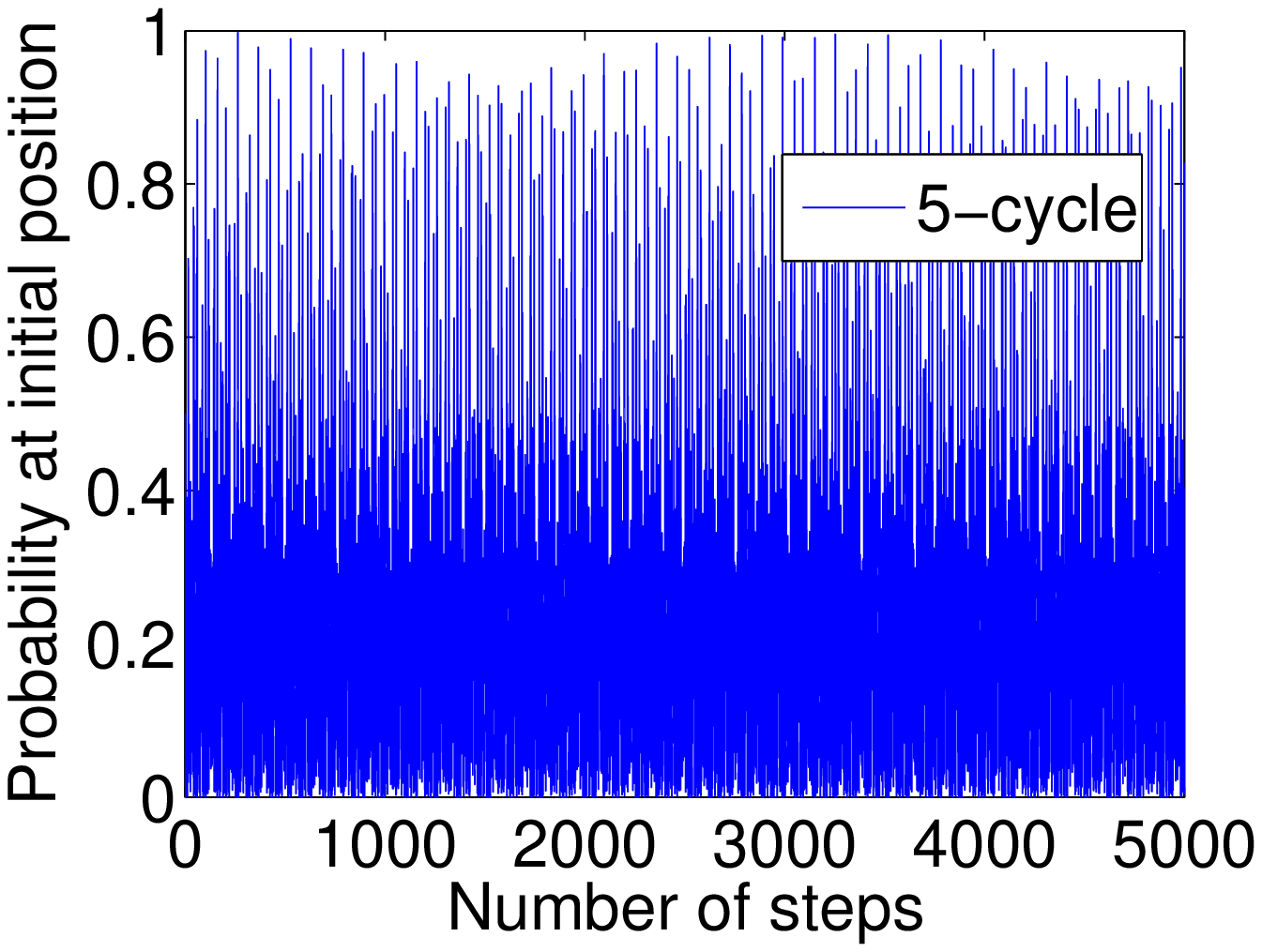, width=7.5cm}
\epsfig{figure=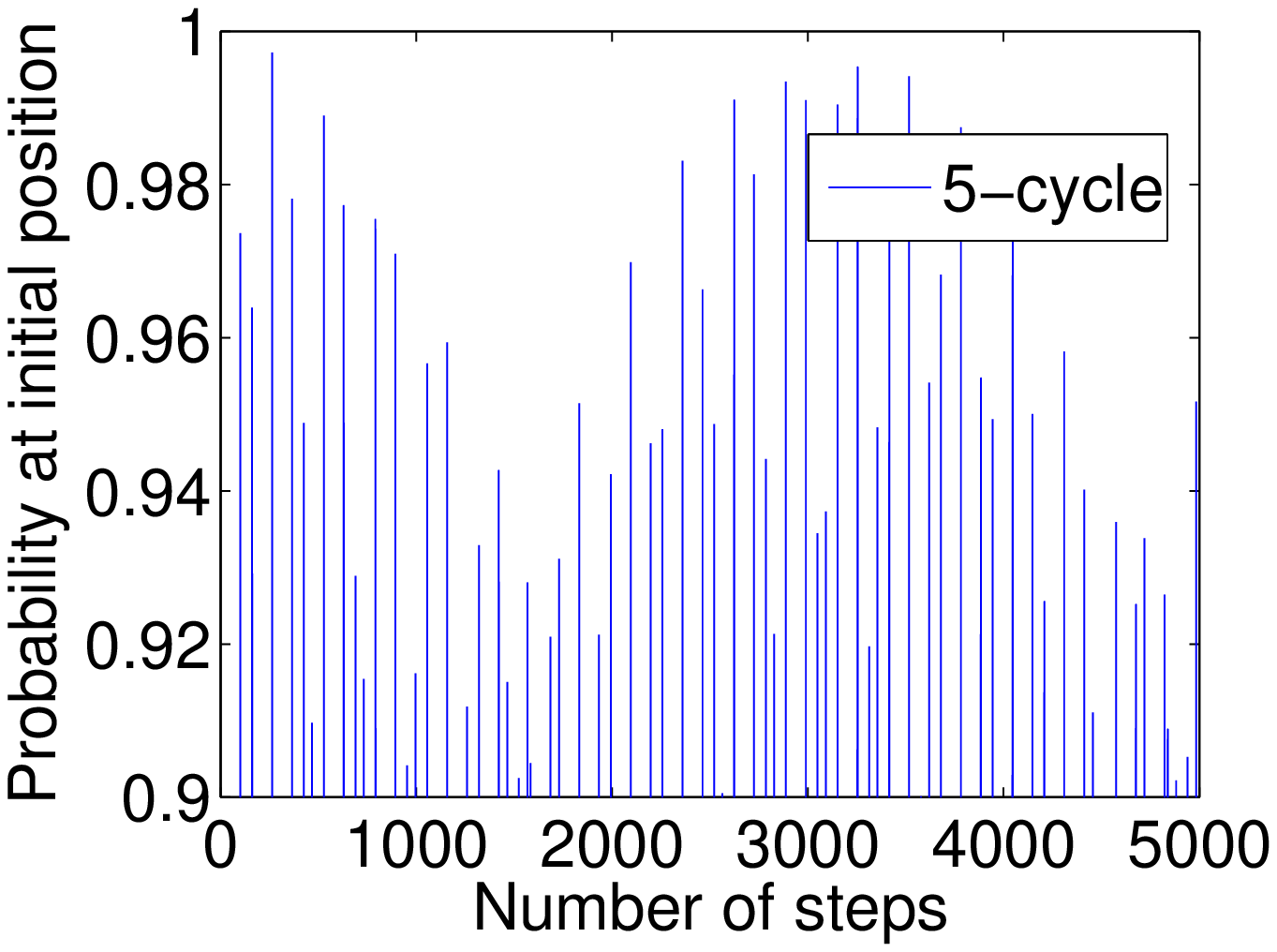, width=7.5cm}\\
~~~~~~~(a)~~~~~~~~~~~~~~~~~~~~~~~~~~~~~~~~~~~~~~~~~~~~~~~~~~~~~~~~~~~~~~~~~(b)
\caption{\label{fig:rec5} Probability at the initial position after different number of steps of QW on $5-$cycle,  (a) is the complete plot whereas (b) is a close up of probability values between 0.9 and 1. At no step in the plot, probability is a unit value.  This numerically shows that the QW on an $n-$ cycle does not recur for n=5.} 
\end{center}
\end{figure}
\end{widetext}
\par
In Fig. (\ref{fig:fracrev1}) we show numerically that the probability of finding the particle at the initial position $x=0$ after different number of steps of QW on $15-$ and $24-$ cycle. The distribution is obtained without time-averaging for up to steps  as large as 5000.  At no time $t =$number of steps, a unit probability value is returned showing that the wave packet evolved using QW fail to revive complete at the initial position. 
\par
The failure of the wave packet to completely revive at initial position and recur can be attributed to the interference effect caused by the mixing of the left and right propagating components of the amplitude. By suppressing the interference effect during the evolution in a closed path, one can get closer to the complete relocalization, revival at initial position $x_0$.  For example, the wave packet can be completely relocalized at $x=0$ on an $n-$cycle after $T=n$ and make the QW recurrent  by choosing  an extreme value of coin parameters $(\xi, \theta, \zeta) =(0^\circ, 0^\circ, 0^\circ)$ in Eq.(\ref{coin}) during the QW evolution. However $\theta = \delta$, $\delta$ being very small and close to $0^\circ$, where the interference effect is minimized, will also return a near complete recurrence on an $n-$cycle. In Fig. (\ref{polyaqw})  the plot of $[1-p_{0}(t)]$ is also shown for a particular case where the QW shows the complete recurrence nature, a walk on an $n-$cycle with $\theta = 0^{\circ}$ and $n=50$. The interference effect is completely suppressed and the quantum recurs after every 50 steps.
\par
From the numerical data for upto 5000 steps  of QW we note that for small $n-$, especially when $n$ is even, the left and right propagating amplitude return back completely to the initial position (origin) before the mixing and repeated interference of the amplitudes takes over at non-initial position in the evolution process. Therefore, for a QW on an $n-$cycle with $n$ being even upto $8$, a complete revival and recurrence of wave packet at initial position $x=0$ is seen Fig. \ref{fig:rev8cycle} \cite{TFM03}. For a QW on particle initially in symmetric superposition state 
\be
\label{qwint}
|\Psi_{0}\rangle = \frac{1}{\sqrt 2} (|0\rangle + i |1\rangle ) \otimes |\psi_{0}\rangle
\ee
with coin operation $C$, Eq. (\ref{qwcoin}) and even $n>8$, due to larger position Hilbert space the interference effect at the non-initial position dominates reducing the recurrence nature of the dynamics. In Fig. \ref{fig:rev10cycle} for QW on $10-$cycle, small deviation from complete recurrence is shown.
\par
For example, we will consider a small odd number, $n =5$, if the position are marked as $x = {0, 1, 2, 3, 4}$,  during the third step of the QW, the left propagating amplitude move from position $2$ to $3$ and the right propagating amplitude move from $3$ to $2$. That is, after the second and third step of quantum walk using shift operator of the form $S^{c}$, Eq. (\ref{eq:condshift1}) and Hadamard operation $H=C_{0, 45^{\circ}, 0}$ as coin operation on a particle initially in superposition state Eq. (\ref{qwint}) takes the form,
\begin{widetext}
\begin{eqnarray}
S^{c}(H \otimes   {\mathbbm  1}) S^{c}|\Psi_{0}\rangle  = \frac{1}{2} \left \{ |0\rangle \otimes |\psi_{3}\rangle +
 \right ( |1\rangle + i |0\rangle \left ) \otimes |\psi_{0}\rangle - i |1\rangle \otimes |\psi_{2}\rangle \right \}
\end{eqnarray}
\begin{eqnarray}
S^{c}(H \otimes   {\mathbbm  1})S^{c}(H \otimes   {\mathbbm  1})S^{c}|\Psi_{0}\rangle = \frac{1}{2 \sqrt{2}} \{ |0\rangle \otimes |\psi_{2}\rangle + \left ( |0\rangle + |1\rangle + i |0\rangle \right ) \otimes |\psi_{4}\rangle  
- \left ( |1\rangle - i|0\rangle + i|1\rangle \right ) \otimes |\psi_{1}\rangle + i |1\rangle \otimes |\psi_{3}\rangle \}.
\end{eqnarray}
\end{widetext}
\par
The left and right propagating amplitude crossover without suppressing the mixing, therefore the constructive interference effect continues to exist at position other than origin during the  evolution. In Fig. (\ref{fig:rec5}) the probability of finding the particle at the initial position is shown. Due to the small size of the Hilbert space, the probability is seen to be close to unity but the closer look reveals that its only a fraction revival.  
\par
We also note that localization effect found in 2D \cite{IKK04} or found in QWs using multi quantum coins to diminish the interference effect \cite{IKS05} can result in increasing the fractional recurrence number $P_{qw}$ on a line and higher dimension. If the particle wave packet is evolved in a {\it position} Hilbert space ${\mathcal H}_P$  with the edges that permits the wave packet to escape, the fractional recurrence nature of the QW does not allow the QW to be completely transient. Therefore, the fractional transient nature of the QW is seen to complement the fractional recurrence nature. 
%===================================
\section{Conclusion} \label{conc}
%===================================
In summary we show that, as long as the wave packet spread in position space interfering, forming small copies of the initial wave packet during the evolution of the QW process, it fails to satisfy the complete recurrence theorem. We show this by analyzing the dynamics of quantum walk process on a line and on an $n-$cycle. We have shown that due to the revival of the fractional wave packet, a fractional recurrence can be seen during the QW process which can be characterized using the quantum P\'olya number or fractional recurrence number.\\
\bc
{\bf Acknowledgement}: 
\ec
I thank Raymond Laflamme for his encouragement and support. I also acknowledge  Mike and Ophelia Lezaridis for the financial support at IQC.

%=================================

\end{document}